\documentclass[traditabstract]{aa} 

\usepackage{graphicx}
\usepackage{wrapfig}
\usepackage{txfonts}
\usepackage{rotating}
\usepackage{natbib}
\usepackage{multirow}
\bibpunct{(}{)}{;}{a}{}{,}
\renewcommand{\sp}{\hspace{1mm}}
\newcommand{\sr}{\hspace{0.5mm}}

\begin{document}
   \title{A low frequency study of PSRs B1133$+$16, B1112$+$50, and
   B0031$-$07}

   \author{R. Karuppusamy\inst{1,2}
          \and
          B. W. Stappers\inst{3,4}
	  \and
	  M. Serylak\inst{4,2}
          }

   \institute{Max-Planck-Institut f\"ur Radioastronomie, Auf dem H\"ugel 69, Bonn, Germany\\
   \email{ramesh@mpifr-bonn.mpg.de}
   \and Sterrenkunde Instituut Anton Pannenkoek, 
              University of Amsterdam, Kruislaan 403, Amsterdam, The Netherlands
      \and   Jodrell Bank Centre for Astrophysics, School of Physics and Astronomy, 
     The University of Manchester, Manchester M13 9PL, UK\\
     \email{Ben.Stappers@manchester.ac.uk}
     \and
     Netherlands Institute for Radio Astronomy (ASTRON), Postbus 2, 7990 AA, Dwingeloo, 
     The Netherlands\\
    \email{serylak@astron.nl}
 }

   \abstract { We use the low frequency (110--180 MHz) capabilities of
     the Westerbork Synthesis Radio Telescope (WSRT) to characterise a
     large collection of single pulses from three low magnetic field
     pulsars.  Using the Pulsar Machine II (PuMa-II) to acquire and
     coherently dedisperse the pulsar signals, we examine whether the
     bright pulses observed in these pulsars are related to the
     classical giant pulse emission. Giant pulses are reported from
     PSR B1112$+$50 and bright pulses from the PSRs B1133$+$16 and
     B0031$-$07. These pulsars also exhibit large intensity
     modulations observed as rapid changes in the single pulse
     intensity. Evidence of global magnetospheric effects is provided
     by our detection of bright double pulses in PSRs B0031$-$07 and
     B1133$+$16. Using the multi-frequency observations, we accurately
     determine the dispersion measures (4.844$\pm$0.002 for B1133$+$16
     and 9.1750$\pm$0.0001 for B1112$+$50), derive the radio emission
     height in PSR B1133$+$16 and report on the properties of subpulse
     drift modes in these pulsars. We also find that these pulsars
     show a much larger intensity modulation at low sky frequencies
     resulting in narrow and bright emissions. }


\titlerunning{Low frequency study of PSRs}
\maketitle

\section{Introduction}
\label{intro}

Most radio pulsars are characterised by a stable average pulse profile
at a given sky frequency. The average profiles are formed by
integrating all pulses emitted by the star over a given time.  The
pulsed emission from the star is observed as individual pulses that
carry detailed information about the physics of the pulsar radio
emission. These pulses can only be directly detected in a fewer than a
third of the known radio pulsars, mainly because of their weak
nature. One of the remarkable features of the single pulse emission is
the occurrence of intense pulses called giant pulses, which are
defined as pulses with energy greater than 10 times the average pulse
energy. Furthermore, the giant pulses are very narrow compared to the
width of average pulsed emission and result in a pulse intensity
distribution described by a power law. For example, in the radio
emission of the young Crab pulsar, the giant-pulse widths are found to
be narrower than 0.4 ns, implying a high brightness temperature of
$\sim$10$^{41}$K \citep{he07}, and the single pulse flux distribution
follows a power law with slope $-$3.3 \citep{lcu+95}. Another example
is the emission of single bright pulses from the young Crab-like
pulsar, PSR B0540$-$69 located in the Large Magellanic Cloud; the
pulsar is visible at radio wavelengths only by virtue of the star's
intense giant pulse emission \citep{jr03}. Similar bright and narrow
pulses were reported in the emission of very old millisecond pulsars
such as PSR B1937+21 \citep{cstt96} and B1821-24
\citep{rj01}. Concerted giant pulse searches in these older and short
period pulsars have revealed more giant pulse sources e.g., PSRs
J1823-3021A and J0218+4232 \citep{kbmo05,kbm+06}. The only common
feature of these pulsars is a strong magnetic field ($B_{LC}
\sim$10$^5$G) at the velocity of light cylinder, which is the limiting
radius where the velocity of the co-rotating plasma and magnetic field
lines is less than the speed of light.

However, some studies propose giant pulse-like emission in
pulsars with a low magnetic field at the light cylinder e.g., $B_{LC}$
in the 4--100 G range. For instance, the bright single pulse emission
from PSRs B0031$-$07, J1752$+$2359, and B1112$+$50 at 40 and 111 MHz
\citep{ke04,ek05,ek03}. \cite{kkg+03} also report bright pulses
from PSR B1133$+$16 at 5 GHz. If these bright pulses are similar to
the classical giant pulses, this questions the requirement of a high
value of $B_{LC}$ to produce giant radio pulse emission. After the
initial discovery, a detailed statistical analysis of the bright pulses
in these low B$_{LC}$ pulsars has not been attempted, unlike the
extensive studies of the giant pulses from the young pulsars and the
millisecond pulsars \citep{lcu+95,kt00,kni07,kbmo06}.

The pulsars in this study also exhibit other interesting single pulse
behaviour: PSRs B1133$+$16, B0031$-$07, and B1112$+$50 show drifting
subpulses \citep{bac73,tmh75,wsw86} and nulling
\citep{hr07,rit76,wsw86,viv95}. PSR B1133$+$16 also shows narrow
emission features called microstructure \citep{han72, cwh90,
  lkwj98}. 

Radio pulsars in general show a negative spectrum \citep{sab86} and a
low frequency spectral turnover \citep{kms+78,mgj+94}. This implies
that pulsars are brighter at low sky frequencies, although the sky
temperature also increases, contributing significantly to the system
temperature. Moreover, microstructure and drifting subpulses can be
more easily studied at low frequencies when sufficient telescope
sensitivity is available \citep{cwh90}. Motivated by the possible
presence of giant pulses and a rich single pulse behaviour in the
pulsars considered here and the availability of the Low Frequency
Front Ends (LFFEs) at the WSRT, we undertook this study in the
110--180 MHz frequency range. The flexible baseband recorder PuMa-II
\citep{kss08} and full coherent dedispersion, permitted a detailed
single pulse study with high sensitivity and relatively high time
resolution. Although these pulsars have been studied at low sky
frequencies before, we are able with our system to observe for much
longer and across a far wider frequency range. Our long observations
are useful in determining the flux distribution and the occurrence
rates of bright pulses. Furthermore, the wide frequency coverage
permits the computation of the pulsar spectra in this band.

The rest of the paper is organised as follows. The observations, the
pulsar fluxes, and spectra are described in Sects. \ref{obs} and
\ref{flux}. We present our revised values of the dispersion measure
(DM) of two pulsars based on the narrow giant pulses in
Sects. \ref{DM} and \ref{bsp}. The single pulse intensity, energy
distributions, and microstructure are the subjects of discussion in
Sects. \ref{spa} and \ref{microstructure}. We proceed with the pulse
drift analysis in Sect. \sp\ref{sss}, which is followed by our
discussion and conclusions.

\section{Observations and data reduction}
\label{obs}

The pulsars were observed with the WSRT on three different days
between June--November 2008 in the 110--180 MHz range (see Table
\ref{ObsDetails}). The design of the WSRT backend systems allows the
selection of a maximum of eight different radio bands within the
frequency range of the front end receiver. Eight 2.5 MHz-wide bands
tuned to various sky frequencies (see Table \ref{ObsDetails}) were
chosen for this study since these bands showed comparatively little
radio frequency interference (RFI). The signals were recorded and
processed using the PuMa-II instrument. When possible, the
observations were carried out in the early hours of the day to reduce
the effect of RFI. Despite our careful choice of the frequency bands
and observing times, these observations were still susceptible to RFI
because of (1) strong in-band RFI in the passband of the LFFE
electronic amplifiers (2) the presence of the automatic gain control
(AGC) in the WSRT's signal chain and (3) the 2-bit system
design. Therefore, the data were subjected to an extensive RFI
cleaning procedure by a combination of automated algorithms and visual
examination.

The WSRT was operated in the tied-array mode by adding signals from 13
telescopes each 25--m in diameter. The analogue signals from the
individual telescopes were digitally sampled at 2-bit resolution and a
fixed rate of 40 MHz. The 2-bit data were then coherently added in a
dedicated hardware unit to produce the 6-bit tied-array data. Coherent
addition improved the signal-to-noise ratio (S/N) by a factor of
$13$. The summed data was then read into PuMa-II as 8-bit numbers,
downsampled by a factor of 8 in realtime, and written to disk. Single
pulses were generated from the coherently dedispersed baseband data
using the dispersion measure in the catalog and based on a polynomial
determined by the TEMPO software \citep{tw89}.

\begin{table*}[ht]
\caption{\label{ObsDetails}Observation details and telescope parameters.}  
\centering
\begin{tabular}{p{0.15\linewidth}ccccccccc}
  \hline\hline
  \noalign{\smallskip}
  Pulsar   & Period  & Date     & Duration &  N$_{pulses}$ & Band \tablefootmark{a}& T$_{sys}$  & N$_{bins}$\tablefootmark{b} & $\Delta t$ \tablefootmark{b} & $S_{min}$\tablefootmark{c}  \\
           &  s      & dd/mm/yy & min      &             &  MHz  &    K      &           &  $\mu$s &   Jy       \\
  \noalign{\smallskip}
  \hline
  \noalign{\smallskip}
  B1133$+$16 \dotfill& 1.187913065936 & 23/11/08 & 280  & $\sim$13701  &110--180 & 400 & 2048 & 809 & 9.0\\
  B0031$-$07\dotfill & 0.942950994559 & 02/11/08 & 360  & $\sim$22758  &110--180 & 400 & 2048 & 580 & 10.7\\
  B1112$+$50 \dotfill& 1.656439759937 & 01/10/08 & 570  & $\sim$14286  &110--180 & 400 & 2048 & 460 & 12.0\\
  \noalign{\smallskip}
  \hline
\end{tabular}
\tablefoot{ 
\\
\tablefoottext{a} {The pulsar signal was recorded from eight 2.5 MHz bands
  centred at 116.75, 130, 139.75, 142.25, 147.50, 156, 163.50, and
  173.75 MHz.}
\\
\tablefoottext{b}{$\Delta t$ is the effective time resolution of the
  pulses with N$_{bins}$ across the pulse period.}
\\
\tablefoottext{c}{ $S_{min}$ is the minimum detectable signal per time sample 
  based on a 30\% aperture efficiency.} 
}
\end{table*}

The open-source DSPSR\footnote{http://dspsr.sourceforge.net/} software
package was used to form a 64-channel software filterbank
simultaneously at the dedispersion stage in each 2.5 MHz band, and the
resulting frequency--time cubes were written to the disk as single
pulses. The original time resolution of 2.5 $\mu s$ and the formation
of a 64-channel synthetic filterbank permitted 2048 samples to be
extracted across the pulse period of the three pulsars observed and
resulted in the final time resolutions in Col. 9 of Table
\ref{ObsDetails}. The large number of frequency channels was later
used to correct for the residual dispersion smearing, since the DMs in
the catalog had either changed or were not sufficiently accurate and
it also aided the removal of narrow band RFI in the signal. Extensive
use of the PSRCHIVE package \citep{hvm04} was made in the computation
of the total intensity, signal-to-noise ratio, and visual inspection
of the single pulses for RFI. Pulse stacks were formed by stacking
individual pulse intensities one above the other according to their
phases. When required, the signal was flux calibrated using the
off-pulse radiometer noise. With the computed fluxes in each band,
various statistical analyses of the data was performed.

\section{Low frequency flux}
\label{flux}

\subsection{Radio frequency interference} 
\label{rfi-clean}

The data was cleaned of interference using various cleaning strategies
for the three purposes of estimation of mean flux density, fluctuation
analysis, and single pulse flux estimates. For the mean pulse flux
computation, the pulses that showed excessive noise, negative drop
outs, and large intensity peaks in the off-pulse region were
removed. The single pulses were combined to create pulse stacks that
were then examined for the second time as time--phase plots; impulsive
interference missed by the automated algorithm was then identified and
removed. The remaining pulses were then averaged over the period of
integration to compute the average pulse profile. The number of pulses
removed in this method was accounted for in the time term of Eq.
\ref{sav} before the profile was flux calibrated. From the calibrated
profiles, the spectral indices were derived.

For the fluctuation analysis, the pulses were first gated to contain
at least three times the width of the average pulse profile at the
lowest observed frequency. The gated pulse stack was then subjected to
the two-pass RFI cleaning procedure explained above. Pulse stacks were
inspected by eye and pulses that showed a saturation in the off-pulse
phases were replaced by zeros. This procedure was repeated until a
visually clean image was seen in the time--phase plane.

The single pulse statistics were computed using only the single pulses
from one of the eight bands observed. RFI was cleaned from the
dedispersed single pulses by visually examining the frequency--phase
images within $30\degr$ of the pulse longitude. For these
comparatively long period pulsars, retaining a narrow phase range of
the pulse period has the advantage of being less sensitive to
impulsive RFI affecting other pulse longitudes. Only pulses that
showed the required dispersion were retained. An interactive plot
utility was developed to examine the frequency--phase images of the
large S/N pulses.

\subsection{Flux calibration}
\label{fcal}

Calibrating pulsar fluxes at low sky frequencies at the WSRT is
difficult for three reasons. First the main contribution to the system
noise $T_{sys}$, is from the sky temperature $T_{sky}$ which in turn
depends on the position of the pulsar on the sky. Second, the aperture
efficiency of the LFFE's at the WSRT is poorly constrained and
thirdly, the presence of RFI introduces errors into the flux
measurements of the calibrators and the pulsars. The system parameters
(telescope gain and system temperature) were derived from continuum
observations of the calibrator source 3C196 made shortly before these
measurements.  From these measurements, the aperture efficiency is
estimated to be $\sim$30\% translating to an effective telescope
collecting area of 1500 $m^2$.  A constant value $T_{sys}$ in the
entire 110--180 MHz is used, since the Galactic synchrotron foreground
varying with frequency as $\nu^{-2.6}$ is almost removed by the
frequency-dependent amplitude gains of the LFFEs. The system
temperatures derived in this manner are shown in
Table. \ref{ObsDetails}. Assuming a frequency scaling of $\nu^{-2.6}$,
the sky temperature estimated from the 408 MHz continuum maps of
\citet{hks+81} in the direction of these pulsars is consistent to
within 20\% of the system temperatures listed in Table
\ref{ObsDetails}.

Once the telescope gain and system temperature had been established,
the off-pulse noise in flux units was computed using the radiometer
equation \citep{dic46} that can be expressed as

\begin{equation}
 \textstyle S_{av} = \frac {\textstyle S_{sys}}{\sqrt{\textstyle N_{p}\cdot B \cdot T_{int}}}\,, 
\label{sav}
\end{equation}

\noindent where $\textstyle S_{sys}=T_{sys}/G$ is the system
equivalent flux density in Jy, $N_p$ the number of polarisations
summed, and $T_{int}$ the equivalent integration time. The pulse
profile was then flux calibrated by multiplying by $\textstyle
S_{av}$. Integrating the fluxes in all phase bins and dividing by the
total number of bins then resulted in the mean continuum flux density
of the pulsar. While this method is not as accurate as the absolute
flux measurements based on noise generators, this is the only option
in a synthesis telescope such as the WSRT for non-imaging
observations; it also yields fluxes comparable to those reported
elsewhere.

In our method of detecting the pulsars, the dedispersed signal was
written to disk as single pulses. Following the formation of total
intensities, diagnostic plots of peak $S/N$ against the phase of the
peaks were made.  Any pulsed emission could then be seen as clustering
of data points near the nominal emission phase and the corresponding
pulses were visually examined. Combining the single pulses in time
after removing the RFI afflicted pulses produced the average emission
profiles. Thus, this method is sensitive to any pulsed emission above
the noise floor.

\begin{figure}[h]
\centering
  \includegraphics[scale=0.95,angle=-90]{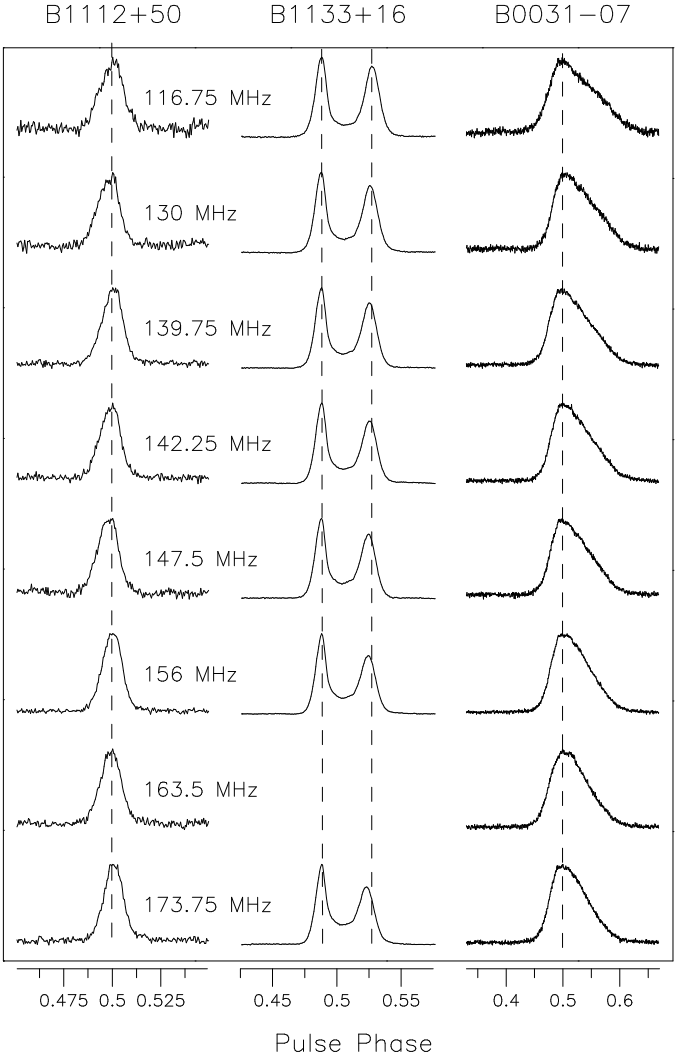}
  \caption{Average pulse profiles of the three pulsars. The profile
    changes with sky frequency is clearly evident. The profile for PSR
    B1133$+$16 at 163.5 MHz is not shown because it was too corrupted
    by RFI. The profiles are shown at 809, 580, 460 $\mu$s time
    resolution for PSRs B1112$+$50, B1133$+$16 and B0031$-$07
    respectively. All profiles are normalised by the maximum flux
    density in the band and are aligned at the phase of maximum
    intensity in each band. Profiles at 116.75 and 130 MHz have
    comparatively more noise because up to $\sim$15\% of pulses were
    removed to reject RFI.}
  \label{profiles} 
\end{figure}

\subsection{Profile evolution}
\label{prof-evol}

The average pulse profiles in all recorded bands for the three
detected pulsars are displayed in Fig. \ref{profiles}. PSR B1112$+$50
displayed 10\% peak widths $W_{10}=$ 35.5--38.8 ms in this frequency
range. This can be compared to $W_{10}=$ 35.0 ms at 405 MHz
\citep{lylg95}. The single component in the average emission of PSR
B1112$+$50 at these frequencies and at 328 MHz \citep{wes06} evolves
to a two-component average emission profile at 1400 MHz \citep{wsw86}.

PSR B0031$-$07 shows $W_{10}=$ 120.2--155.1 ms, while \citet{lylg95}
report a value of 104.5 ms at 405 MHz. The trailing edge in the
average profile of PSR B0031$-$07 shows a gradual decrease in
steepness with frequency. We rule out scattering in the ISM as a
possible cause of this decrease because that would manifest itself as
pulse broadening by a factor of $\sim$5 in the 110-180 MHz band
considered here, which is clearly not seen (e.g Figure
\ref{profiles}). However, this change in slope is indicative of the
emergence of a second emission component at even lower frequencies e.g
at 40 and 62 MHz \citep{ikl+93} and represents almost the opposite
behaviour of the average emission profile in PSR B1112$+$50.  Another
contribution to the slope of the trailing part of the average emission
profile in PSR B0031$-$07 could be the vertically drifting subpulses,
which is discussed in Sect. \ref{sss}. The characteristic age of PSRs
B1112$+$50 and B0031$-$07 is 10 Myr and 36 Myr, respectively. The
general expectation of a narrower pulse profile for older pulsars is
not valid with respect to these pulsars, as the impact angle of the
observer's line of sight to the emission beams from the pulsars may
not be equal.

PSR B1133$+$16 displays a profile width of $W_{10}=44.0 \mbox{--}
48.1$ ms compared to 41.8 ms at 405 MHz \citep{lylg95}. The component
separation of the average profile decreases with frequency and this
phenomenon was successfully explained by the radius-to-frequency
mapping in the pulsar magnetosphere \citep{cor78}. The average profile
also shows a double emission component, with a pronounced bridge
emission that rises from 14\% to 25\% of the peak flux density at our
lowest and highest frequencies, respectively. The second component in
the average emission profile also reduces from 90\% to 70\% of the
leading component intensity with increasing frequency. These
variations clearly probe different parts of the emission
beam. \citet{now95} assumed a model in which the pulsar beam consists
of two concentric emission cones to explain the observed two-component
emission profile and the increasing bridge emission with
frequency. However, this model is unable to explain why multiple
emission components are not visible at higher
frequencies. Polarisation observations of this pulsar over a range of
frequencies would offer further insights into the emission
geometry. As discussed later in this work and by \citet{knn+82}, the
emission height of various radio frequencies in the magnetosphere of
the pulsar can be derived using the frequency dependence of the
component separation. Our analysis shows that a $\nu^{-0.30\pm0.02}$
relation is valid for the component separation in the frequency range
considered here and is slightly steeper than the $\nu^{-0.24}$ and
$\nu^{-0.26}$ scaling laws derived by \citet{bac72} and \citet{srw75},
respectively.

\subsection{Radio spectrum}
\label{spec}

\begin{figure}[h]
  \includegraphics[angle=-90]{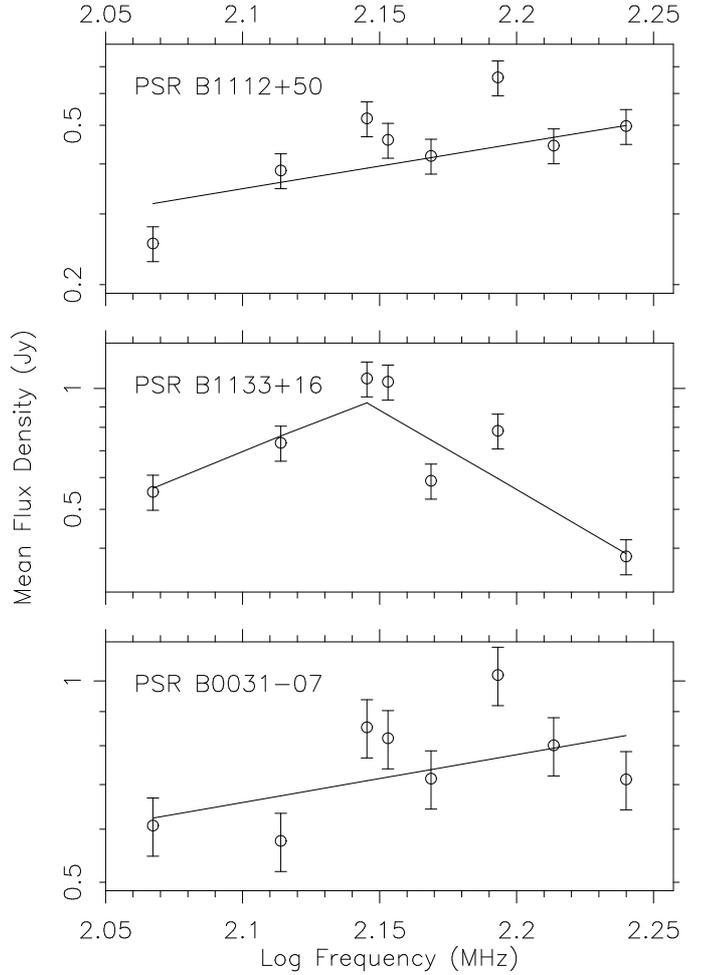}
  \caption{Spectra for three pulsars in the frequency range 100--200
    MHz. Errors bars correspond to a 10\% uncertainty in the pulsar
    flux estimation. The best fit power law curves are also
    displayed. For PSR B1133$+$16, a broken power-law provides a
    closer fit to the measured fluxes. The flux in the RFI-affected
    band centred at 163.5 MHz is not included for PSR B1133$+$16.}
  \label{sindex} 
\end{figure}

The observations presented here span a reasonably wide frequency range
and we use the flux of the pulsar in each of the recorded bands to
compute the pulsar spectra. The pulsar flux used here is the mean
continuum flux density of these sources, computed from the
flux-calibrated pulse profiles. The functional form of the radio
spectrum of the pulsars was assumed to follow $S(\nu) \propto
\nu^\alpha$, where $S(\nu)$ is the pulsar flux at frequency $\nu$ and
$\alpha$ is the spectral index. The parameter $\alpha$ was estimated
using a least squares fitting procedure and the result is displayed in
Fig. \ref{sindex} and Table \ref{sindex-tbl} for the three pulsars. To
account for the effect of RFI and possible errors in telescope
parameters, a 10\% uncertainty in the estimation of the pulsar flux
density in each band is considered, which dominates other errors in
S/N estimates because all pulsars are strongly detected.

We now consider the effects of scintillation on the estimation of the
spectral indices. The length of these observations are between 5 and 9
hours (see Table. \ref{ObsDetails}) and may be susceptible to
scintillation effects on these timescales. Diffractive scintillation
does not influence our results here as we now demonstrate. The
frequency--phase plots displayed in Fig. \ref{all-gps} shows that the
intensity modulation induced by this type of scintillation is confined
to a few channels. To help validate this, the diffractive
scintillation bandwidth and time were derived from values published in
other work (see Table. \ref{sindex-tbl}), using $\Delta \nu_{DISS}
\propto \nu^{4.4}$ and $\Delta t_{DISS} \propto \nu^{1.2}$ scaling
laws. From the foregoing discussion, it is clear that the observed and
derived diffractive scintillation bandwidth is narrow for these
pulsars at low frequencies and that we average over many ``scintiles''
with the wide band used here. The effects of the refractive
scintillation should also be considered for these long observations,
and this timescale is derived as outlined in \citet{sc90a} and listed
in Table. \ref{sindex-tbl}. Hence, it is clear that the diffractive
scintillation does not affect the results derived here. The refractive
scintillation may affect the these long observations, but are
accounted for in the 10\% uncertainty in flux estimation considered
here. The derived timescale $T_r$ shows that refractive scintillation,
if there is any, affects only the spectral index estimation for PSR
B1112$+$50.

\begin{table*}[!ht]
\caption{\label{sindex-tbl}Measured and derived parameters of three pulsars.}
\centering
  \begin{tabular}{p{0.15\linewidth} c c c c c c c}
    \hline\hline
    \noalign{\smallskip}
    Pulsar  & $S$ \tablefootmark{a}&  $\alpha_1$  & $\alpha_2$ & $\Delta\nu_{DISS}$ \tablefootmark{c}& $T_d$ \tablefootmark{c}& $T_r$\tablefootmark{c} & Reference\\
            & $m$Jy     &              &            & kHz             & secs. & days & \\  
    \noalign{\smallskip}
    \hline
    \noalign{\smallskip}
    B1112$+$50  \dotfill & 460\sp$\pm$\sp46  & 1.12$\pm$0.56 & - & 77 -- 447 & 115 -- 187 & 2.6 -- 1.1 & 1\\
    B1133$+$16 \tablefootmark{b} \dotfill  & 1030\sp$\pm$\sp51  & 2.33$\pm$2.55 & -3.81$\pm$2.24 & 7 -- 40 & 48 -- 77 & 11.8 -- 4.9 & 2\\ 
    B0031$-$07\dotfill   & 821\sp$\pm$\sp41  & 0.71$\pm$0.45 & - & 3 -- 19 & 212 -- 343 & 109.1 -- 45.2 & 3\\
     \noalign{\smallskip}
    \hline
  \end{tabular}

  \tablefoot{
\\     
\tablefoottext{a}{The flux density quoted here is the value measured at 142.25 MHz.}
\\
\tablefoottext{b}{A broken power-law spectrum results in a better fit for this pulsar.} 
\\
\tablefoottext{c}{Scintillation parameters are derived from --
(1) \citealt{cr98}; (2) \citealt{kkg+03}; (3) \citealt{sc90a}}
}

\end{table*}

PSR B1112$+$50 shows a slightly positive spectral index and we find a
value of $\alpha = 1.12 \pm0.46$. Even though this pulsar is the
weaker of the three pulsars, it is detected with high significance, as
is clearly evident from Fig. \ref{profiles}. The somewhat poor fit to
the pulsar spectra in these observations is not caused by fluxes
corrugated by diffractive scintillation for the reasons explained
above and may have its origins in the residual RFI and/or refractive
scintillation. The results presented here imply that the spectra turns
over at a frequency higher than 180 MHz.

For PSR B1133$+$16, the middle panel of Fig. \ref{sindex} can be
compared to Fig. 4e in \citet{deshrad92}, which extends down to 34.5
MHz. Their figure can be interpreted as indicating a spectral turnover
in our frequency range. With this consideration, the spectrum was
modelled by a two component power-law defined as

\begin{equation}
S(\nu) = \left\{
\begin{array}{l l}
  C_1 \nu^{\alpha_1} & \qquad \mbox{    if   $\nu < \nu_b$ }\\
  C_2 \nu^{\alpha_2} & \qquad \mbox{    if $\nu \ge \nu_b$},\\ 
\end{array} \right.
\label{brkplaw}
\end{equation}

\noindent where, $\nu_b$ is the break frequency, $C_1$ is a constant,
and normalisation yields $C_2=C_1 \nu_b^{\alpha_1 - \alpha_2}$. The
values of break frequency of 139.75 MHz with spectral indices
$\alpha_1=2.33\pm2.55$ and $\alpha_2= - 3.81\pm2.24$ were derived from
the best fits to the data. The effect of scintillation gives rise to
``scintiles'' with typical widths of $\le 40$ KHz, which tend to
average out within each 2.5 MHz band and do not contribute to the
intensity variations in the 110-180 MHz frequency range. Combined with
the 75\% error in 400 mJy at 151 MHz \citep{sie73} and 900 mJy at 111
MHz (Figure 3 in \citealt{mgj+94}), we conclude that our results are
consistent with those reported previously.

We find a relatively flat spectrum for PSR B0031$-$07 with an index of
$\alpha=0.71\pm0.45$. As for the other two pulsars, B0031$-$07 has
partially resolved scintillation (see Fig. \ref{all-gps}) at these
frequencies that do not contribute to the estimated fluxes.  For sky
frequencies $\nu\le 1$ GHz, \citet{mgj+94} find that the spectral
index of this pulsar is $\alpha=-1.2\pm0.26$. Considering the two data
points below $\nu= 100$ MHz and the error limits in \citet{mgj+94},
the spectral index derived here is consistent with their work.  The
trend seen in the spectra of this pulsar implies that the spectra
turns over at a frequency higher than 180 MHz.

\section{Bright single pulses}
\label{bsp}

\begin{figure*}[htbp]
  \center
  \includegraphics[angle=-90]{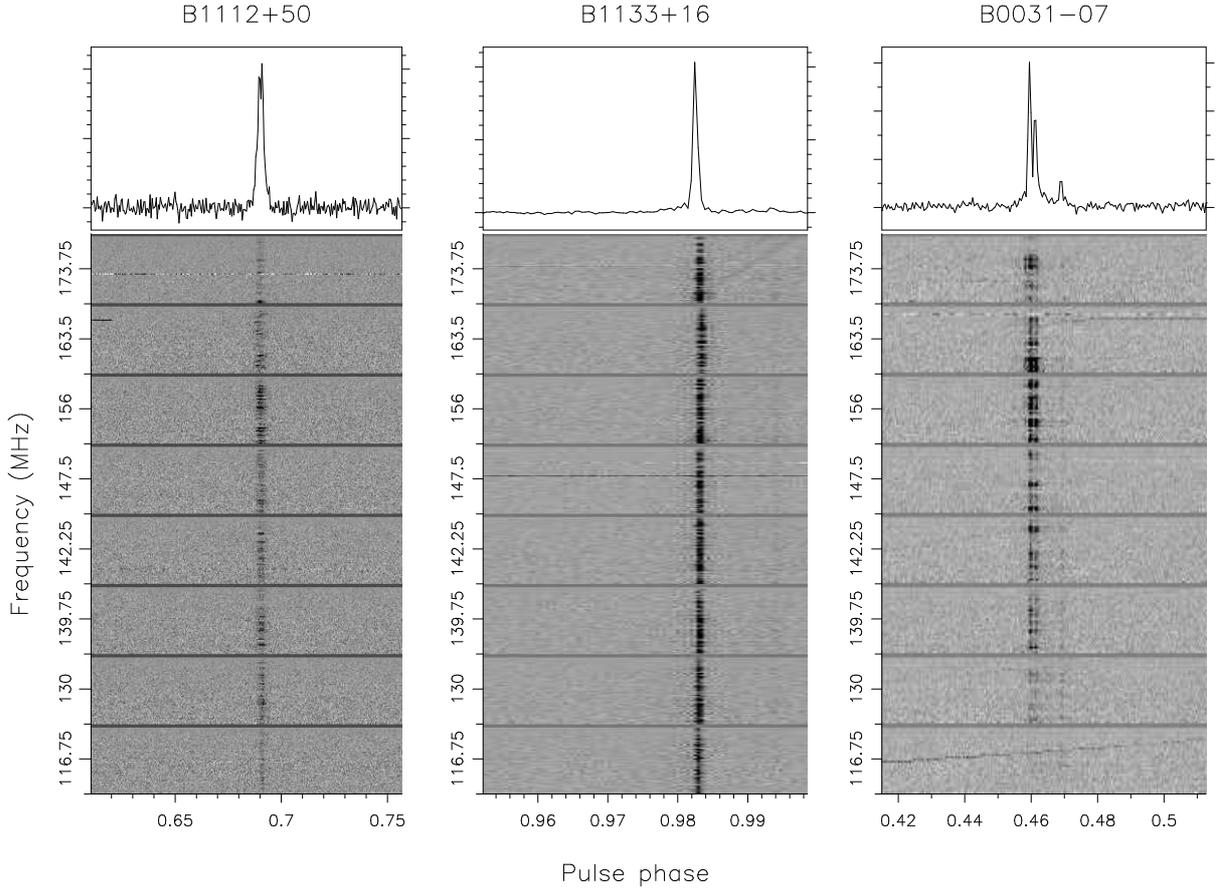}
  \caption{Narrow bright pulses from the three pulsars detected at all
    eight sky frequencies. The total intensity displayed in the top
    panels is computed by summing the signal in the 8 bands. The lower
    panels show a dedispersed pulse as a frequency--phase plot. The
    abscissa in these plots correspond to the pulse phase at 116.75
    MHz. Some residual RFI is visible as low-level intensity
    modulating the entire pulse phase in the frequency-phase plots of
    all three pulsars. }
  \label{all-gps} 
\end{figure*}

Several bright pulses were detected from all three pulsars and
examples are shown in Fig. \ref{all-gps}. These pulses are very narrow
and often less than 1 ms in width. In general, giant pulses are
characterised by their widths, the phase of the pulses with respect to
the average emission profile, and the pulse energy \citep{kni07}. The
classical giant pulses are also found to be broadband, for example in
B1937+21 $\Delta \nu/\nu \sim$ 0.2 \citep{ps03}, where $\nu$ and
$\Delta \nu$ are the observation frequency and bandwidth. In this
study, we find similar bright, broadband bursts from the pulsars
studied, although they are not as energetic as the classical giant
pulse emitters.

\begin{table}[!h]
\caption{\label{tb-tbl}Brightness temperature of single pulses.}

\centering
  \begin{tabular}{p{0.35\linewidth}cc}
    \hline\hline
    \noalign{\smallskip}
    Pulsar      & d  & $T_B$  \\
                & kpc & K   \\
    \noalign{\smallskip}
    \hline
    \noalign{\smallskip}
    B1112$+$50 \dotfill  & 0.32\tablefootmark{a} & $3.51\times 10^{26}$ \\
    B1133$+$16 \dotfill  & 1.06\tablefootmark{b} & $1.93\times 10^{27}$ \\
    B0031$-$07\dotfill   & 0.35\tablefootmark{c} & $1.32\times 10^{27}$ \\
    \noalign{\smallskip}
    \hline
  \end{tabular}

\tablefoot{ 
\\
\tablefoottext{a}{ DM distance from \citealt{tc93}.}
\\
\tablefoottext{b}{ Parallax distance from \citealt{cbv+09}.}
\\
\tablefoottext{c}{ Parallax distance from \citealt{bbgt02}. }
 }
\end{table}

The implied brightness temperature of the single pulses can be
computed from (pp. 79; \citealt{lk05}),

\begin{equation}
 T_B = \bigg(\frac{S_{peak}}{2\pi k_B}\bigg) \bigg(\frac {\nu \Delta t}{d}\bigg)^{-2} \,,
\label{tb}
\end{equation}

\noindent where $S_{peak}$ is the peak pulse flux, $k_B$ is the
Boltzmann's constant, $d$ is the distance to the pulsar, $\nu$ the sky
frequency and $\Delta t$ is the pulse width. Using $\Delta t$
values from Table \ref{ObsDetails} and the known distances for these
pulsars, the computed brightness temperatures are listed in
Table \ref{tb-tbl}. 

While the brightness temperatures of these pulses are several orders
of magnitude lower than the $10^{35}$ -- $10^{40}$ K seen in typical
giant pulses from the Crab pulsar or PSR B9137$+$21, the large values
of $T_B$ do point to a coherent emission mechanism for these
pulses. From the frequency--phase plots in Fig. \ref{all-gps}, these
bright pulses clearly represent broadband emission, with $\Delta \nu
/\nu =$ 0.3, and can be compared to the value of 0.1 and 0.2 for the
Crab pulsar and PSR B1937+21 at 1400 MHz. The scintiles in all three
pulsars are also visible in this figure, though they are more evident
in PSRs B1133$+$16 and B0031$-$07.  Therefore these bright
  pulses are not due to the occasional focussing effects brought about
  by scintillation. The similarities between the bright pulses found
  here and the classical giant pulses are indicative of high energy
  counterparts as in the Crab pulsar and PSR B1937+21.

\section{Dispersion measure variations}
\label{DM}

Pulsar dispersion measures are normally estimated from multi-frequency
multi-epoch timing observations \citep{pw92,bhh93}. However, in some
pulsars the absence of an appropriate fiducial point in the average
emission profile introduces errors in the estimated DM
values. \citet{agmk05} used cross-correlation of the pulse profiles at
two different frequencies of observation to partially overcome this
problem. These authors also point out that an alternative means of
estimating the DM is to use multi-frequency simultaneous observation
of the pulsar, and use a fiducial point in the average emission
profile to align the profiles at the two frequencies. Our detection of
single, narrow, and bright pulses at low sky frequencies provides in
an excellent reference to align the pulses and hence compute the DM
very accurately for these low DM pulsars. The frequency resolution of
64 channels across each 2.5 MHz band gives a total of 512 channels in
the 110--180 MHz range, which permits the accurate determination of
the quadratic dispersion curve. We generally started with the catalog
DM, which was varied until a peak S/N in the combined pulse profile
was attained. The frequency-phase plots were then visually inspected
for the pulse alignment using the new value of DM. To compute the
change in the DM at the epoch of our observations, the dispersion
smearing caused by the use of an incorrect DM within a single
frequency channel was considered negligible.

From their multi-year timing data, \citet{hfs+04} find DM$=$4.864
cm$^{-3}\sr$pc for PSR B1133$+$16 corresponding to the epoch MJD 46407
and also determine
d(DM)$/$d$t\approx-0.0008\sr$cm$^{-3}\sr$pc$\sr$yr$^{-1}$. Our
analysis shows that the DM has changed to 4.844$\pm$0.002
cm$^{-3}\sr$pc at epoch MJD 54793 from our reference value of 4.871
cm$^{-3}\sr$pc at epoch MJD 41665. The error estimate comes from the
DM uncertainty caused by the misalignment of a broadband pulse by one
phase bin across the 110--180 MHz band. The total DM change,
$\Delta$DM, found here is consistent with the predictions of
\citet{hfs+04}. The somewhat high yearly rate of change in the DM of
this pulsar probably arises from its large proper motion $\sim$630
km$\cdot$s$^{-1}$. The $\Delta$DM$=0.0265$ cm$^{-3}\sr$pc corresponds
to a 0.5\% change in DM and failing to correct for this change leads
to the peak flux of the bright pulses being underestimated by
$\sim$3\%.

For PSR B1112$+$50, we find that the DM increased from 9.16
cm$^{-3}\sr$pc at an epoch MJD 50899 to 9.1750$\pm$0.0001
cm$^{-3}\sr$pc at epoch of MJD 54795 resulting in
$\Delta$DM$=$0.015$\sr$cm$^{-3}\sr$pc. The error in our DM value is
derived from the pulse misalignment by one phase bin in a 2.5 MHz band
at 147.5 MHz. We again refer to \citet{hfs+04} for DM estimates of
this pulsar, where they find DM$=$9.195$\sr$cm$^{-3}\sr$pc and a
yearly rate of
d(DM)$/$d$t=-0.004\sr$cm$^{-3}\sr$pc$\sr$yr$^{-1}$. These values are
inconsistent with the rate of
$-$0.0014$\sr$cm$^{-3}\sr$pc$\sr$yr$^{-1}$ derived from our
analysis. One reason for this discrepancy could be the difference
between the high frequency observations at 408--1600 MHz used in the
work of \citet{hfs+04} and the much lower frequencies used in our
study which are more sensitive to DM changes. The second reason could
be that a single value of d(DM)$/$d$t$ is insufficient implying that
higher order terms of the DM time derivative may be necessary to
accurately model the DM variations.

We do not detect a change in the DM of PSR B0031$-$07 from the
reference DM value of 10.89 cm$^{-3}$pc. The narrow and bright single
pulses have no discernible errors when the signal is dedispersed with
this DM. However, the pulsar's DM of 10.89 cm$^{-3}$pc is inconsistent
with the value of 11.29 cm$^{-3}$pc at this epoch expected from the
$-0.007$ cm$^{-3}$ pc yr$^{-1}$ listed in \citet{hfs+04}. Moreover, in
the studies of \citet{tml93}, \citet{ke04}, and \citet{sms+07}, the
authors use a DM of $10.89$ cm$^{-3}$pc, which implies that the DM of
the pulsar has not changed since 1993. Therefore both the DM value and
the rate of DM change for this pulsar in \citet{hfs+04} may be
incorrect.

\section {Single pulse analyses}
\label{spa}

\begin{figure}[h]
  \includegraphics[angle=-90]{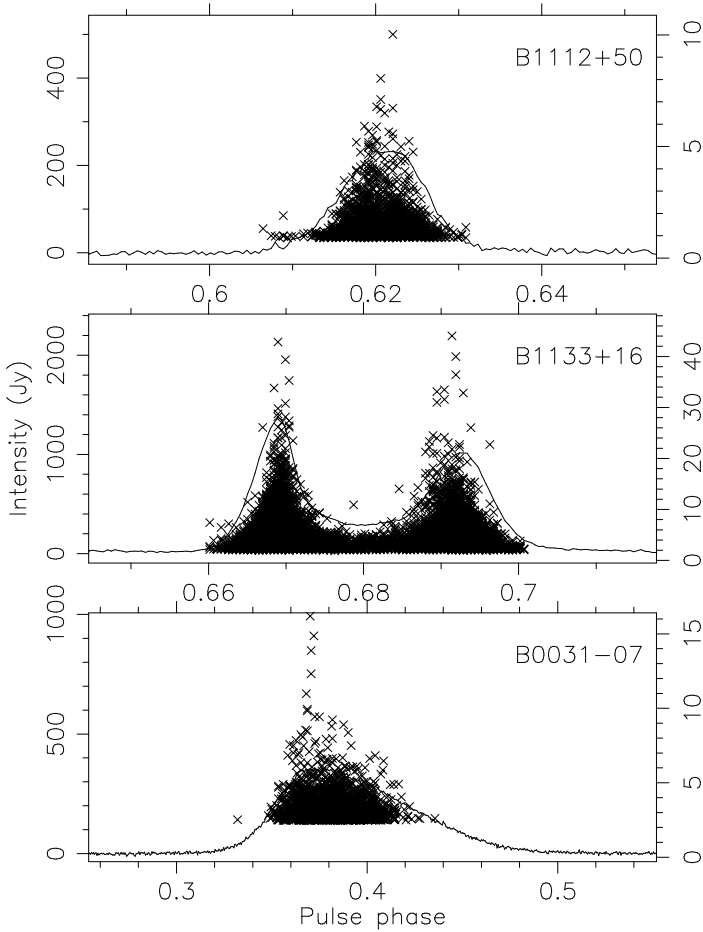}
  \caption{Single bright pulses and the integrated pulse profiles at
    156 MHz for the three pulsars. The flux scale on the left ordinate
    corresponds to the peak flux of the single pulses and the scale on
    the right ordinate axis corresponds to the average pulse
    profile. The phase displayed here is different from that of
    Fig. \ref{all-gps} due to the signal being dedispered to a
    different central frequency. Peak flux densities of the single
    pulses are shown as crosses. Only pulses $\ge 20 \sigma$ are shown
    in these plots which represent a low flux cut-off in the fluxes of
    single pulses.}
  \label{profile-gps} 
\end{figure}

Our long observations resulted in the detection of a large number of
pulses and are shown in Table \ref{ObsDetails}. Robust statistics can
be obtained by computing the distribution of single pulse energies $E$
and peak intensities $S_p$. A scatter plot of the peak fluxes and the
average emission profile for the three pulsars are shown in Fig. 
\ref{profile-gps}. The 2.5 MHz band at 156 MHz was chosen out of the
eight bands for further study because it showed comparatively less
RFI. We note that after an exhaustive RFI search and removal, all
eight bands show similar trends.

\begin{figure}[h]
  \includegraphics[angle=-90]{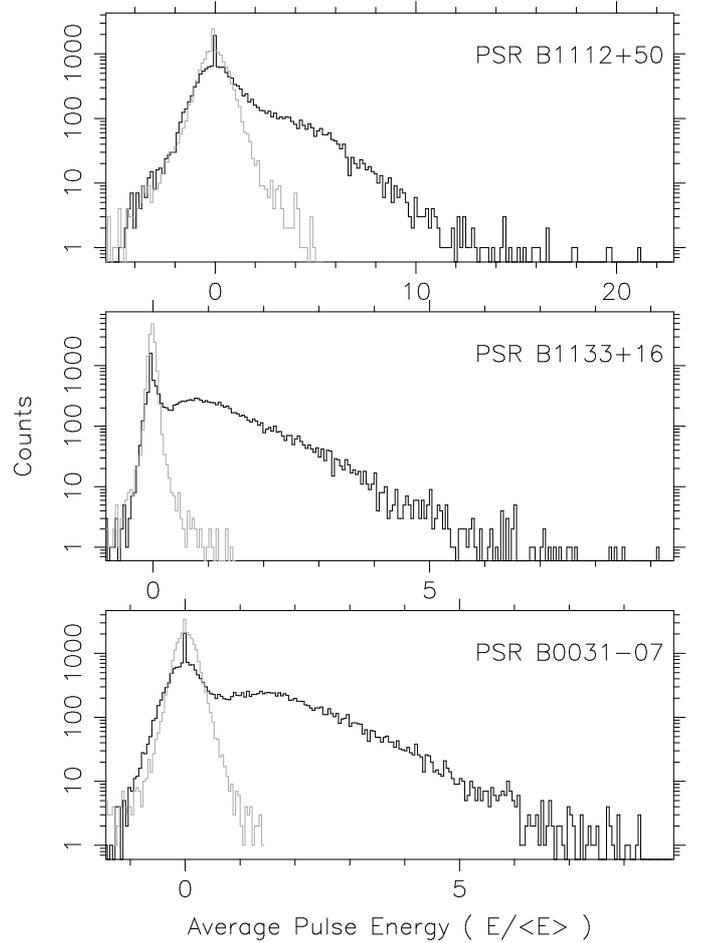}
  \caption{Pulse energy distribution for the three pulsars. The
    abscissa shows pulse energy normalised to the mean pulse
    energy. Dark and light grey lines correspond to the on-pulse and
    off-pulse energy distributions. The clear peak at zero energy is
    introduced by the RFI affected pulses replaced by zeros.}
  \label{penergy-fig} 
\end{figure}

\begin{figure}[h]
  \includegraphics[angle=-90]{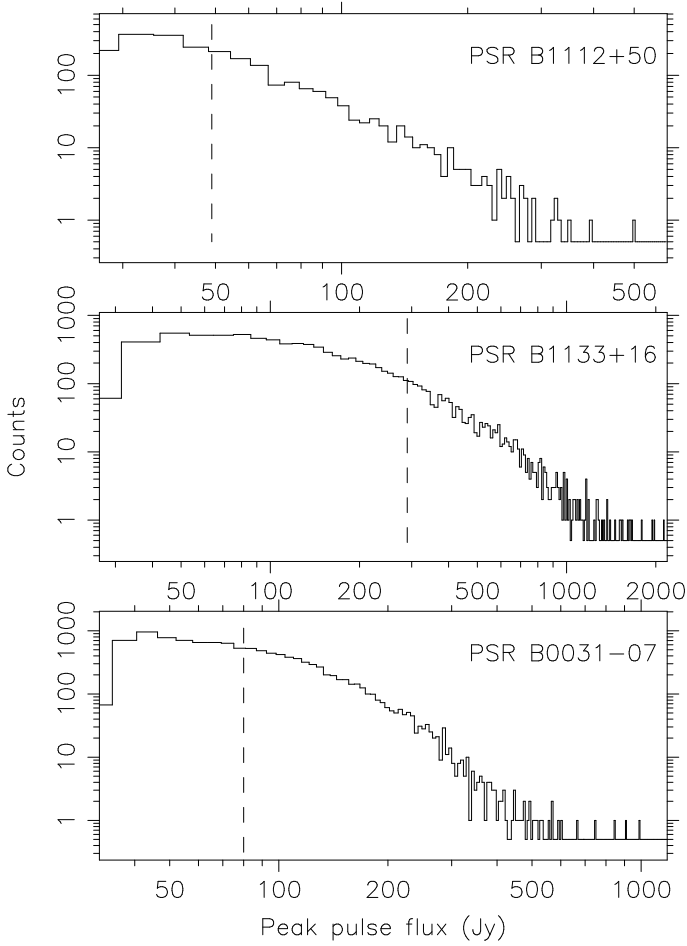}
  \caption{Plot of the peak flux distribution of the single pulses for
  the three pulsars. The dashed line shows $10\times S_{av,peak}$ for
  each pulsar.}
  \label{peakflux-fig} 
\end{figure}

The presence of narrow bright pulses was evident when the data was
examined visually in the RFI cleaning stage. These unusual emission
entities should be evident in the single pulse energy and flux
distributions. For example, giant pulses in the emission of the Crab
pulsar should result in a power law in the pulse flux distribution
\citep{lcu+95}. For the three pulsars considered here, the
distribution of the pulse flux and energy are shown in Figs.
\ref{penergy-fig} and \ref{peakflux-fig}, respectively.  The pulse
energy was computed by summing the emission in a window of width equal
to the average emission profile. The single-pulse peak flux was
computed using the product of the peak $S/N$ and $S_{av}$
(Eq. \ref{sav}).

\subsection {PSR B1112$+$50}

The top panel of Fig. \ref{profile-gps} shows that this pulsar emits
bright pulses. Some of these pulses show considerable structure, with
narrow subpulses separated by a few milliseconds. These pulses have a
slight preference to the trailing phase of the average pulse profile,
although a clear phase relationship is not evident. This is similar to
the giant pulse-like emission from this pulsar reported by
\citet{ek03}.  Fig. \ref{penergy-fig} shows the pulse energy
distribution and the bright pulses with energies greater than $10
\times \langle E \rangle$ comprise the trailing part of the
distribution. An indication of the null emission in this pulsar is
visible from the slight turn-over in the pulse energy distribution
near $3 \times \langle E \rangle$, which is similar to the offpulse
energy distribution. Our data is consistent with the $\ge$ 60\%
nulling fraction at 1412 MHz \citep{wsw86} and at 408 MHz
\citep{rit76}. If the null emissions are accounted for with this
nulling fraction, then $\langle E \rangle$ increases by 60\%. This
causes a decrease in the energy of bright pulses in the 10--13 $\times
\langle E \rangle$ range. The pulse flux distribution also displays a
shallow power-law like tail in the peak flux distribution (top panel
in Fig.  \ref{peakflux-fig}) produced by the occasional bright pulses
in this pulsar.

\subsection {PSR B1133$+$16}

\citet{kkg+03} reported bright pulses from this pulsar at a frequency
of 5 GHz but not at 1.4 GHz.  We detected bright, narrow pulses in all
eight bands in the range 110-180 MHz with peak fluxes greater than 10
times the peak of average flux density as seen in the middle panel of
Fig. \ref{peakflux-fig}. The pulse energy distribution shown in the
middle panel of Fig. \ref{penergy-fig} can be qualitatively
approximated by a power law. The distribution does not exhibit pulses
with energies $>$$10\times \langle E \rangle$, which is the normal
working definition of giant pulses \citep{cstt96}. From the middle
panel in Fig. \ref{profile-gps}, a preference for the bright pulses to
fall on the inner edge of the two components is evident.  Some of
these bright pulses contain two distinct and narrow subpulses. Bright
pulses from this pulsar at 327 MHz were also noted by
\citet{hr07}. The turn-over in the pulse energy distribution $\sim
$1.0$\times\langle E \rangle$ is indicative of a bimodal distribution,
and hence of nulling. The pulsar shows a $\sim$20\% null fraction at
327 MHz \citep{hr07}, but at our frequency we are unable to identify
the null distribution because we were limited by the system
sensitivity.

\subsection {PSR B0031$-$07}

\citet{ke04} reported giant pulses from PSR B0031$-$07. Our
observations reveal that the bright and spiky pulses do not satisfy
the 10$\times \langle E \rangle$ criterion but are easily greater than
$10 \times S_{av,peak}$, as seen in Figs. \ref{profile-gps} and
\ref{peakflux-fig}. Fig. \ref{profile-gps} also reveals the emission
of bright pulses on the leading edge of the average pulse profile. The
work of \citet{ek05} (see their Table 1) suggests that the bright
pulses in this pulsar have a steep spectral index. We also draw the
reader's attention to Fig. 4 of \citet{ke04}, where the authours
report the detection of ``double giants'' in this pulsar at 40 and 111
MHz. These pulses were detected in our observations, as seen in the
rightmost panel of Fig. \ref{all-gps}. These bright subpulses with two
distinct peaks show structure spanning a few milliseconds implying
that they are not microstructure in the emission. As with PSR
B1133$+$16, a turn-over in the pulsar energy distribution
$\sim$1.5$\times \langle E \rangle$ is indicative of null emission.

\section{Microstructure}
\label{microstructure}

\begin{figure}[h]
  \includegraphics[scale=0.85,angle=-90]{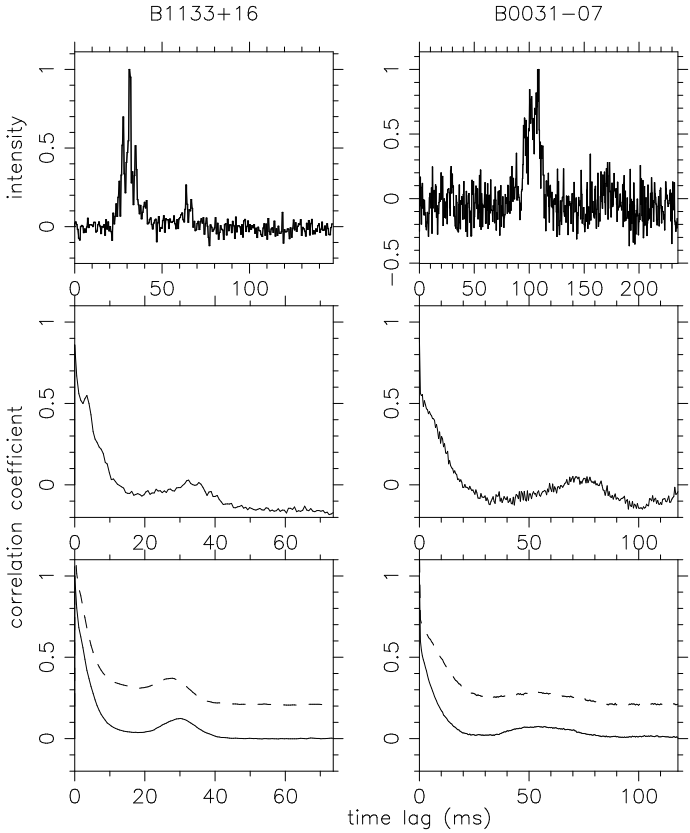}
  \caption{Intensity autocorrelation functions (ACF) of pulses with
    structure for two pulsars. The uppermost panels present example
    pulses. The autocorrelations displayed in the middle panel
    correspond to the pulses above them. The two bottom panels display
    the averaged ACFs for 120 and 132 pulses for PSRs B0031$-$07 and
    B1133$+$16 respectively. The averaged ACF's for two bands centred
    on 116.75 MHz and 173.25 MHz (dashed line) are shown. }
  \label{micro-fig} 
\end{figure}

Several single pulses from both PSR B1133$+$16 and B0031$-$07 show
pulse structures considerably narrower than the width of the average
pulsed emission. The presence of pulses with microstructure became
evident when several thousand pulses were examined in the process of
cleaning RFI from the data. To quantify the microstructure, the
autocorrelation function (ACF) \citep{cwh90} of the single pulses and
their averages were computed. The discrete ACF is defined by

\begin{equation}
\label{acf}
R(\tau) = \displaystyle\sum_{t=0}^{nbin}I(t) \cdot I(t+\tau),
\end{equation}

\noindent where $I(t)$ is the intensity of the pulsar. Fig. 
\ref{micro-fig} shows $R(\tau)$ for a single pulse and the averaged
ACF of several pulses.

For PSR B1112$+$50, the single pulses were quite narrow, devoid of any
feature that can be recognised as microstructure. Therefore, this
pulsar is not considered any further in this section. The separation
between the two main emission components can be clearly seen for
B1133$+$16 and is the feature at correlation lag $\sim$30 ms. The ACF
for the pulse shown in the uppermost panel exhibits the periodicity of
microstructure (i.e., the local maxima near correlation lag 5 ms in
the middle panel of Fig. \ref{micro-fig}). It is known that PSR
B1133$+$16 shows structure down to $\sim$8$\mu$s and a characteristic
intensity fluctuation of the order of $\sim$575 -- 663 $\mu$s in the
111--196 MHz range \citep{han72}. Our final resolution of 580 $\mu s$
was sufficient neither to resolve narrower structures nor to permit us
to estimate the persistent quasi-periodic intensity variations in the
averaged ACF.

In the case of PSR B0031$-$07, single pulses display narrow bursty
features. However, the single pulse ACF in the right middle panel of
Fig. \ref{micro-fig} shows no characteristic variations implying that
the pulse consists of non-periodic microstructure. These nonperiodic
features were also found in PSR B0950+08 \citep{lkwj98}. The average
subpulse separation is seen in the averaged ACFs and is visible at
$\sim$55 $ms$ for PSR B0031$-$07. Expressed in pulse longitude, this
is $20\degr$, which is also the value of $P_2$ found in Sect.
\ref{sss}. For the average ACF's displayed in the two lowest panels of
Fig. \ref{micro-fig}, the pulses chosen are detected simultaneously in
the two bands. The slight shift of local maxima to the left cannot be
measured with high significance, implying that the subpulse separation
decreases with frequency as can be seen by comparing the subpulse
properties at 328 and 1380 MHz in \citet{wes06} and \citet{wse07}.
The slight break at $\sim$10 ms in the ACF of the single pulse
represents the width of microstructure in the pulsar. The reason for
the lack of a prominent characteristic timescale in the microstructure
of PSR B0031$-$07 may be our coarse time resolution.

\section{Fluctuation spectra}
\label{sss}

Since the three pulsars under discussion display bright and bursty
emission, it is natural to expect considerable pulse-to-pulse
variation. In some pulsars (eg. PSRs B0809+74, B1944+17), drifting and
persistent microstructure were detected leading \citet{cwh90} to
suggest that these two phenomena are correlated.  Moreover, PSRs
B1133$+$16 and B0031$-$07 show drifting subpulses at other
frequencies. To ascertain the drifting behaviour at these frequencies,
we undertook sub-pulse drift analysis.

\begin{figure*}[htbp]
\includegraphics[scale=1.1]{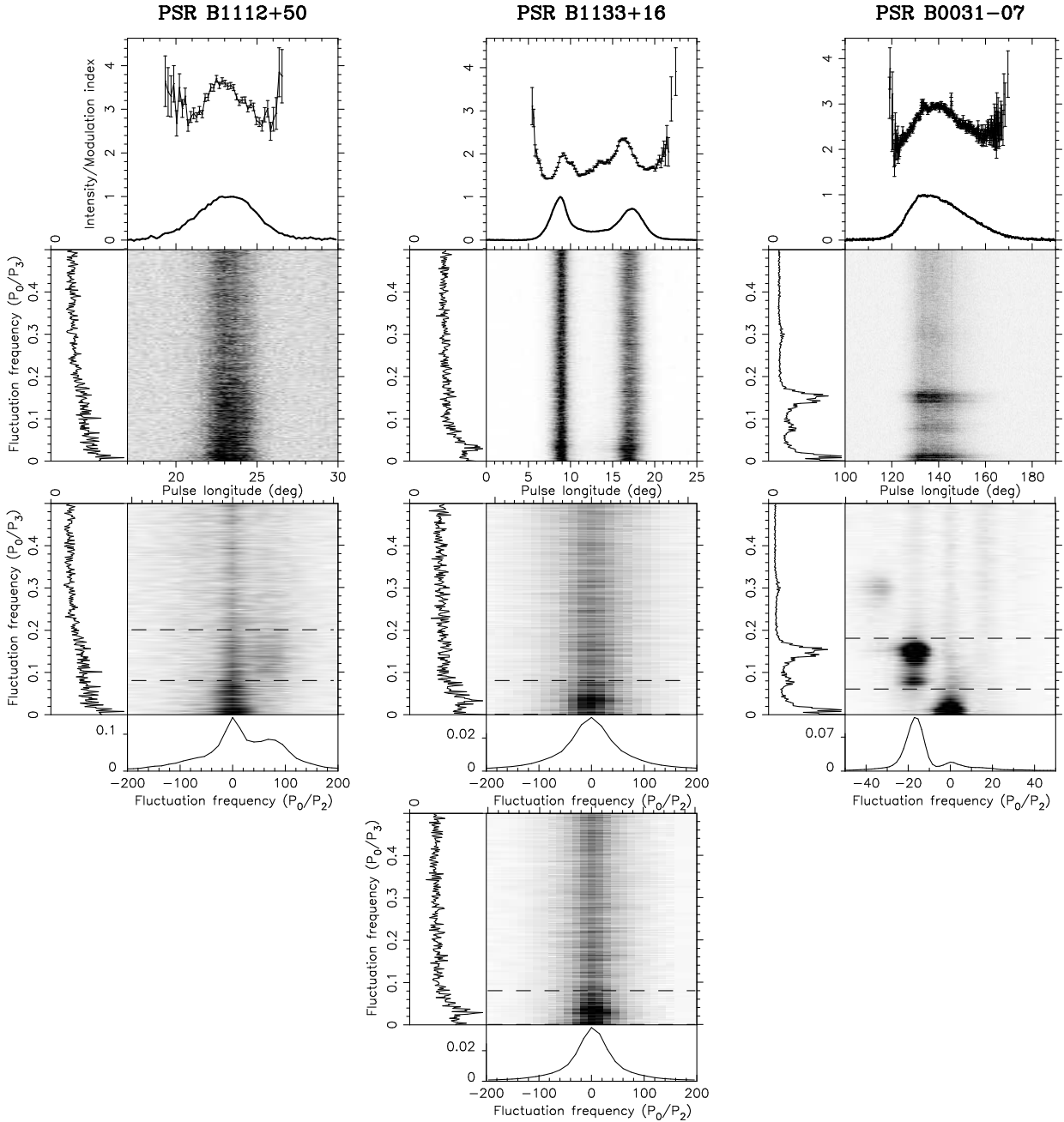} 
\caption{Plots of
  the fluctuation analysis for the PSRs B1112$+$50, B1133$+$16, and
  B0031$-$07. The average pulse profile, and the longitude-resolved
  modulation index (LRMI) are shown in the uppermost panels. The
  panels directly below the average pulse profile present the
  longitude-resolved fluctuation spectrum (LRFS). The two-dimensional
  fluctuation spectra (2DFS) are shown in the third and fourth
  rows. For PSR B1133$+$16, the 2DFS plots are shown separately
  corresponding to the two components of the average pulse
  profile. The panels on the left and bottom of the LRFS and 2DFS
  images contain the vertically and horizontally integrated spectra,
  respectively.}  \label{2ds}
\end{figure*}

To perform the analysis presented in this section, the pulse stacks
were subjected to a pulse drift analysis based on the
longitude-resolved modulation index (LRMI), longitude-resolved
fluctuation spectrum (LRFS), and two-dimensional fluctuation spectrum
(2DFS) \citep{es02,wes06}. The sliding two-dimensional fluctuation
spectrum (S2DFS) was also computed to identify the onset of the
various drift modes \citep{ssw09}. The resulting plots from the LRMI,
LRFS, and 2DFS analysis are shown in Fig. \ref{2ds}. In this plot,
only LRMI values that are detected with a significance of more than
$6\sigma$ are shown (top panel; Fig. \ref{2ds}). In the LRFS and 2DFS
plots, drifting subpulses produce regions of enhanced brightness
called ``features''. The vertical drift rate $P_{3}$ and the
horizontal separation $P_{2}$ of the subpulses are defined to be the
centroid of a rectangular region in the 2DFS containing the features
(Eq. 6; \citealt{wes06}). The sign of $P_{2}$ denotes the direction of
the drift, i.e., whether subpulses drift towards the leading
(negative) or trailing (positive) edge of the pulsar's average
profile, respectively. In what follows, the S2DFS analysis results are
discussed only for PSR B0031$-$07. The analysis of PSRs B1133$+$16 and
B1112$+$50 using the S2DFS technique provided results that are
comparable to those obtained from the 2DFS method alone.

The results presented in this section are for the 2.5 MHz band centred
on 156 MHz, as in the previous section. In the RFI cleaning procedure
discussed in Sect. \ref{rfi-clean}, fewer than 2\% of the pulses were
replaced by zero, and therefore the effect is considered small enough
to be neglected. The drift analysis was performed on all eight bands,
and the results presented here are representative of the other
bands. Within the 110--180 MHz range, there are no noteworthy changes
in drift features. The modulation index was also in general agreement
with the results found in the band reported here but are higher than
the values reported in previous high frequency studies (see Table
\ref{drift-tbl}).

\begin{table}[h]
\centering
\caption{\label{drift-tbl}Drift analysis results for three pulsars.}
\begin{tabular}{p{0.22\linewidth}rrrrr}
\hline\hline
\noalign{\smallskip}
\multirow{2}{*}{Pulsar}  & $P_{3}$  & $P_{2}$   & \multicolumn{3}{c}{$m$} \\
\noalign{\smallskip}
\cline{4-6}
\noalign{\smallskip}
                       & $[P_{0}]$ & [deg]      & 21cm & 92cm & 192cm \\
\noalign{\smallskip}
\hline
\noalign{\smallskip}
B1112$+$50 \dotfill      & $6 \pm 3$            & $5^{+4}_{-2}$         & 1.5  & 2.1  & 3.5 \\
\noalign{\smallskip}
\hline
\noalign{\smallskip}
\multirow{2}{*}{B1133$+$16\tablefootmark{a} . . .}   & $30 \pm 5$     &  $-$  & 1.4  & 0.8  & 2.2 \\
\noalign{\smallskip}
                        & $34^{+5}_{-3}$       &  $-$    &      &      &     \\
\noalign{\smallskip}
\hline
\noalign{\smallskip}
\multirow{2}{*}{B0031$-$07\tablefootmark{b} . . .} & $12.2^{+0.2}_{-0.1}$ & $-21^{+2}_{-1}$       & 1.2  & 1.4  & 3.1 \\
\noalign{\smallskip}
                                      & $6.7 \pm 0.1$        & $-21.6^{+0.5}_{-0.2}$ &      &      &     \\
\noalign{\smallskip}
\hline
\end{tabular}
\tablefoot{ Vertical and horizontal separations of the subpulses are
  listed in Cols. 2 and 3, respectively. The value of modulation index
  $m$ is the minimum in the longitude-resolved modulation index and is
  shown for three different wavelengths in the last three columns. The
  values in Col. 4 are from \citet{wes06}, values in Col. 5 from
  \citet{wse07}, and the results in the last column are from this
  work.
\\
\tablefoottext{a}{ Values presented for the leading and trailing component.}
\\
\tablefoottext{b}{ Values presented for the drift mode ``A'' and ``B''.}
 }

\end{table}

\subsection{PSR B1112$+$50}

This pulsar shows a broad, non-specific vertical drift feature as seen
in the panels on the left in Fig. \ref{2ds}. \citet{wsw86} reported
the presence of three different drift modes in this pulsar and that
the dominant mode in the trailing part of the average pulse profile
showed a periodicity of $P_3\sim 6 P_0$. At 328 MHz, \citet{wse07}
report that $P_2 = 40^{-10}_{+20}$ degrees and a $P_3 = 9 \pm 5
P_{0}$. We detect $P_2 = 5^{+4}_{-2}$ degrees and subpulse modulation
of $P_3 = 6 \pm 3 P_{0}$. We also note a greater amount of power in
the entire range of $P_{3}$ centred around $P_{0}/P_{2} = 0$. While
the $P_3$ found here is consistent with values reported elsewhere
within the errors, the discrepancy between the $P_{2}$ results
reported by \citet{wse07} and our own can be explained by the higher
S/N of individual pulses in our observations. This results in a more
prominent feature in the spectrum and less convolution with the higher
power in the entire range of $P_{3}$ centered around $P_{0}/P_{2} =
0$.

It is possible that only the prominent drift mode active at 1412 MHz
(mode 2 of \citealt{wsw86}) is also active at this frequency. The
support for this comes from the similarity of the average profile at
our frequency to the average profile computed from 193 pulses
containing drift mode 2 in \citet{wsw86}. In their work, the profile
computed from 176 pulses in mode 1 displays a prominent double
peak. From the LRMI plot, the modulation index $m \sim 3.5$ is quite
high at this frequency. Combined with the values reported elsewhere
(see Table \ref{drift-tbl}), this indicates that the pulse-to-pulse
intensity variations increase with decreasing frequency. This again
suggests that the large amplitude pulses are more common in this
pulsar at lower sky frequencies. The LRFS shows an excess near
$P_0/P_3 \sim 0$ owing to the large nulling fraction of $\sim$60\% in
the emission of this pulsar \citep{rit76}, which can be also seen in
Fig. \ref{penergy-fig}. However, this can only be indicative of
nulling because some contribution to the excess seen comes from the
zeros introduced by the RFI cleaning procedure. Therefore, these plots
cannot be used to measure the pulsar's nulling fraction.

\subsection{PSR B1133$+$16}

The results of the analysis presented here confirms the presence of
the long-period feature $P_3 \sim 30$ at this frequency, which was
previously reported at frequencies greater than 300 MHz \citep{wes06,
  hr07}. In Fig. \ref{2ds} (middle panel), the horizontal component of
the drift feature, $P_2$ is very weak and consistent with no ordered
horizontal separation of the subpulses. The feature near $P_0/P_3 =
0.18$ is present in both components of the pulse profile and the LRMI
plot shows the presence of a significant modulation in the bridge
emission region. We find $P_{3} = 30 \pm 5 P_{0}$ for the leading
component and $P_{3} = 34^{+5}_{-3} P_{0}$ for the trailing
component. Apart from the aforementioned $P_3$ feature, the LRFS shows
a non-ordered feature that modulates the emission throughout the
entire stretch of the analysed pulse sequence. This is also evident
from the elevated baseline of the collapsed spectra shown in the side
panels of Fig. \ref{2ds}. The excess power seen near $P_0/P_3 = 0$ is
due to the combination of null emission and the pulses replaced by
zeros in the process of cleaning RFI, though as in PSR B1112+50 this
cannot be measured reliably from this analysis.

\subsection{PSR B0031$-$07}

The three different drift modes in this pulsar were first reported by
\citet{htt70} of which two are clearly visible in our analysis
(Fig. \ref{2ds}, right-hand panel). The very low frequency feature at
$P_0/P_3 \sim$0.02 comes from the nulls present in the pulse
stack. The nulls are a combination of the null emission and the zeros
introduced by RFI cleaning procedure. We detect this feature in all
the eight observed bands. The feature seen at $P_0/P_3 = 0.15$ denotes
the prominent drift mode ``B'' that introduces harmonics and hence
appears as the feature seen at $P_0/P_3 =$ 0.3. The other feature seen
at $P_0/P_3 = 0.08$ is the drift mode ``A'', which can be identified
with regular upward drifting bands in the pulse stack. The drift mode
``C'' is not seen in the 2DFS spectrum. We used the S2DFS method to
analyze the occurrence rate of different drift modes. While modes
``A'' and ``B'' were clearly detected in the S2DFS output, mode ``C''
remained undetected. The mode ``C'' has a low occurrence rate and is
active only for $\sim$2.6\% of the observation time at 327 MHz
\citep{vj97}. However, \citet{smk05} detect this mode marginally at
328 MHz but not at 4.85 GHz, indicating a steeper spectral index for
the drift mode. Therefore, other than susceptibility to RFI and lower
sensitivity, it is unclear why this mode is not detected at these
frequencies.

\section{Discussion}

The long observations reported in this paper provide good sensitivity
to persistent weak features in the average emission of the pulsars. We
were thus able to test the hypothesis of interpulse emission in PSR
B1112$+$50 proposed by \citet{wsw86} at 1400 MHz. Based on the outer
gap emission model then proposed, these authors recommend the use of
observations at low sky frequencies to detect the inter pulse, as the
model predicts large fanbeams. Our deep exposures at low sky
frequencies shows no evidence of inter-pulse emission with an upper
limit of 3.5 mJy at 156 MHz (see Fig. \ref{profiles}), which is $<$
0.6\% of the main pulse peak flux. The non-detection of an interpulse
in this pulsar is unsurprising given that only $\sim$1.5\% of the 1847
known pulsars possess an interpulse emission component
\citep{wj08}. Examination of polarimetric profiles at 400 and 608 MHz
from the Jodrell Bank Observatory archives shows that the position
angle swing is too shallow, which would result in a large impact angle
to the observed radio beam.  Our observations and the indication from
past polarisation studies shows that if the neutron star has a dipolar
magnetic field, PSR B1112$+$50 is not an orthogonal
rotator. Consequently, the alternate magnetic pole never sweeps across
our line of sight.

The pulsar spectral indices found in this study indicate a turn-over
near 180 MHz. \citet{mic78} modelled the pulsar emission as consisting
of a coherent part and an incoherent synchrotron component. This model
also predicts a spectral index of $-8/3$ and a spectral turn-over
below 100 MHz based on the finite energy of the injected particles,
and that the radiation is dominated by incoherent emission
mechanisms. However, this contrasts with the presence of bright pulses
and clear single pulses near the putative turnover frequency implying
that the radiation is still of a coherent nature. The simulations of
\citet{gk92}, which assume that the pulsar radiation is produced by
emission cones, clearly illustrate that the estimated intensity is
very sensitive to the viewing geometry. If the radiation originates
quite high in the magnetosphere (as at low sky frequencies), it is
possible that the viewing geometry changes quite rapidly. This in turn
can give rise to the observed pulse flux near the spectral turn-over
resulting in spectra that does not vary smoothly. Detailed
polarisation studies at these frequencies can be used to test whether
the observed frequency dependence of the pulse flux has its origins in
the viewing geometry, although the pulse shape is quite stable. In
summary, past studies have been inconclusive on the value of spectral
index for PSRs B1133$+$16 and B0031$-$07 at low sky frequencies. The
results presented in this work suggest a spectral turn-over near
$\sim$140 MHz for PSR B1133$+$16 and above 180 MHz for PSRs B0031$-$07
and B1112$+$50.

As pointed out by other authors, the discrepancies in pulsar DMs stem
from the difficulty in aligning multi-frequency pulse profiles and the
choice of a fiducial point in the average profile
\citep{hfs+04,amg07}. The narrow bright pulses at low sky frequencies
in the pulsars studied in this work have permitted a very accurate
estimation of the DM. We note that scatter broadening of the emitted
pulses can limit the accuracy of the DMs estimated by both other
methods and ourselves. This can be another source of the DM
discrepancies. Because of the broad band nature of the pulsar emission
mechanism, we cannot rule out the claim that some pulsars may have
slightly different values of DM in different parts of the frequency
spectrum \citep{knn+82}. However, with the S/N in this work we do not
measure any deviation from the $\nu^2$ dependence of the interstellar
medium contribution to dispersion in the 110--180 MHz
band. Simultaneous broad-band observations with the existing
instruments and LOFAR can address these outstanding questions
effectively.

The analyses of the single pulse energy distributions revealed that
only PSR B1112$+$50 exhibits pulses of energy $\ge 10 \times \langle E
\rangle$. The presence of null emissions in these pulsars leads to an
underestimate of the average pulse energy. Therefore, this is an
important consideration when interpreting pulse energy distributions
for pulsars that exhibit the nulling phenomena. The width and
intensity of the bright pulses in this study are quite similar to
those of the normal giant pulses suggesting that they are caused by
similar emission mechanisms. Sources of classical giant pulses display
a phase correlation between the giant pulses and the high energy
non-thermal X-ray emission, as in the Crab pulsar \citep{lcu+95}, PSR
B1937+21 \citep{chk+03}, and in other millisecond pulsars
\citep{kbmo06}. However, these are energetic pulsars with large values
of $B_{LC}$. It is known that PSR B1133$+$16 shows X-ray emission
\citep{kpg06}, which the authors claim contains both thermal and
non-thermal components. With their coarse time resolution of 3.2 s and
low photon counts, these authors were unable to detect any X-ray
pulsations. However, the non-thermal high energy emission may be
related to the bright pulses observed in our work. For PSRs B1112$+$50
and B0031$-$07, a deep search in the archival X-Ray data might reveal
any correlation that might exist. The bright double-pulses reported in
this work and by \citet{ke04} are indicative of a global
magnetospheric phenomena, despite the different physical locations of
the emitting regions. However, it could also be that drifting
occasionally brings these two regions into our line of sight resulting
in the double pulses. Detailed analysis of these double pulses will
provide insight into the phenomena that might be active over the whole
magnetosphere.

The preference of the bright pulse emission to occur at the phase
coincident with the peak in the average emission profile is a feature
common to all pulsars in this study. This argues that these pulses are
preferentially emitted further away from the last open magnetic line
in the magnetosphere, in contrast to pulsars such as PSR B1937+21
\citep{kt00}. The bright pulses in B0031$-$07 are more frequent and
bright at even lower sky frequencies when we compare our results in
the 110--180 MHz range with those found by \citet{ek05} at 40
MHz. This is also valid for the other two pulsars studied here --
bright pulses are more common in these pulsars at lower
frequencies. This finding contrasts with the models of \citet{pet04}
in which the bright and narrow pulses at high frequencies were
explained by inverse Compton scattering of low frequency radio
photons. Simultaneous multi-wavelength radio observations might help
to resolve this issue.

In the pulsar models proposed by \citet{rs75}, the radio waves are
emitted tangentially to the diverging magnetic field lines by the
relativistic electron--positron pairs. In this model, the width of the
observed average profile is then defined by the last open magnetic
field lines. Furthermore, the model predicts that the emitted radio
frequencies $\nu$ follows a $\nu \propto r^{-3/2}$ relation, where $r$
is the radial distance from the centre of the star. This leads to the
lower frequencies being emitted higher in the magnetosphere than the
high frequency radio waves, giving rise to a radius-to-frequency
mapping (RFM). Our observations show a decrease in the component
separation in the average pulse profile of PSR B1133$+$16 with
frequency, which scales as $\nu^{-0.3}$. After correcting for the
residual $DM$ smearing, the retardation and abberation time delays
predicted by the RFM model can be measured from the 116 and 173 MHz
average profiles. Using a method similar to \citet{cor78}, we find a
lower limit of $5.6 \times 10^5$ m for the height of 116 MHz radio
emission. Depending on the methods used, previously measured emission
heights were $< 10^{5.0}$--$10^{5.8}$m and $< 10^{4.7}$--$10^{5.8}$m
for the radio emission in the 40--1400 MHz range
\citep{cor78,mw80}. The upper limit in these works corresponds to the
40 MHz radiation emitted at radius $r < 10^{5.8} = 6.3\times
10^5$m. Therefore, the derived emission height of $5.6 \times 10^5$ m
for the 116 MHz radiation is consistent with the former results. For
the other two pulsars, a reduction in the profile width is visible in
Fig. \ref{profiles}. The lack of a reliable fiducial point in the
average profiles of these two pulsars prevents a robust estimate being
made of the retardation and abberation time delays, hence estimate of
emission heights.

The drift analysis revealed the large modulation index in all three
pulsars at this frequency, which are similar to that of the Crab
pulsar ($m=$ 5 to 8) as found by \citet{wes06} and \citet{wse07} at
1380 and 382 MHz, respectively. Combined with the values in their
work, the modulation index of the pulsars considered here increases
with decreasing frequency indicating the prevalence of bright and
spiky emission. From their observations in the 341--4850 MHz range,
\citet{kkg+03} report giant pulses in PSR B1133$+$16 at 4850 MHz, with
a majority of them occurring at the phase of the leading component of
the average emission profile. They also report individual pulses at
1412 MHz that can be distinguished clearly from the noise, but have
less than 10 times the average pulse energy. Furthermore, these pulses
were contemporaneous with the giant pulse emission at 4850 MHz. If the
pulses we detect are related to those found at 4850 MHz, then this
raises the question of why the mechanism responsible for the bright
pulses is not active at 1412 MHz. A possible reason is the unusual
spectral behaviour of these bright pulses. Alternatively, the average
emission may have a greater contribution from the normal pulses at
1412 MHz than at 4850 MHz or at 116 MHz. Simultaneous observations
spanning even wider frequencies than \citet{kkg+03} could help us to
answer this question.

We note that the slopes at the edges of the average emission profiles
seen in PSRs B1112$+$50 and B0031$-$07 are indicative of the direction
of the dominant drift mode. These pulsars were also identified as
drifters by \citet{wse07}. From their work, it can also be seen that
several pulsars exhibiting coherent drifting indicate that one of the
leading or trailing edges of the average emission profiles is steeper
than the other (e.g B2012$+$51, B2310$+$42, B2016$+$28 etc.).

\section{Conclusions}

The main conclusion that can be drawn from the pulse energy
distributions for PSRs B1133$+$16 and B0031$-$07 is that they show a
power-law like form in the frequency range observed here. PSR
B1112$+$50 displays a pulse energy distribution with a somewhat
steeper power law as a result of several pulses with energy
$\ge$10$\times \langle E \rangle$. Even though the pulses detected are
much narrower than the average pulse profile, these characteristics
alone do not qualify these pulsars as giant pulse emitters in the
classical sense. However, a single pulse analysis for these pulsars
does reveal large modulation indices, suggesting that they are similar
to the ``spiky'' emission observed in the low frequency observations
of B0656+14 \citep{wwsr06}. Extending this argument, this emission may
be related to the RRAT-like emissions \citep{wsrw06}.

In this work, the spectrum of three pulsars in the 110--180 MHz range
have been derived. PSR B1133$+$16 has been found to exhibit a spectral
break in this range, while the spectra of PSRs B1112$+$50 and
B0031$-$07 display spectra peak at frequencies higher than 180
MHz. From our detections of narrow and bright pulses, we reported DM
changes for PSRs B1112$+$50 and B1133$+$16, but no change in the DM of
PSR B0031$-$07.  We have detected microstructure in PSRs B1133$+$16
and B0031$-$07, although with our sensitivity we are unable to measure
the characteristic intensity variation timescales. In a drift analysis
of the single pulses in PSR B0031$-$07, two of the three drift modes
were detected. The modulation indices computed from the pulse stacks
of all three pulsars have values larger than those at higher
frequencies. This shows the presence of strong pulse to pulse
intensity variations and probably arises from the combination of
drifting subpulses, single bright pulses and emission nulls. While it
remains unclear whether these pulsars emit giant pulses, the emission
does change considerably as one moves to lower radio frequencies.

\begin{acknowledgements}
 We acknowledge the use of European Pulsar Data network, the ATNF
 pulsar catalogue for this work and the SAO/NASA Astronomical Data
 System maintained by Harvard-Smithsonian Center for Astrophysics. The
 WSRT is operated by ASTRON/NWO. We thank the observers in setting up
 the observations. The PuMa-II instrument and RK were funded by
 Nederlands Onderzoekschool Voor Astronomie (NOVA). MS is supported by
 the EU Framework 6 Marie Curie Early Stage Training programme under
 contract number MEST-CT-2005-19669 (ESTRELA). RK thanks Priya for
 help in certain aspects of data analysis. We thank the anonymous
 referee for comments which improved this manuscript.
\end{acknowledgements}

\bibliographystyle{aa}

\end{document}